\documentclass[manuscript]{aastex62}
\usepackage{amsmath}
\newcommand{\Alfven}{$\rm Alfv\acute{e}n~$}
\shorttitle{Turbulent Proton Heating Rate in the Solar Wind}
\shortauthors{Sasikumar Raja et al.}
\begin{document}

\title{Turbulent Proton Heating Rate in the Solar Wind from $5$ to $45~R_{\odot}$}

\correspondingauthor{K. Sasikumar Raja}
\email{sasikumar.raja@iiap.res.in ; sasikumarraja@gmail.com}

\author[0000-0002-1192-1804]{K. Sasikumar Raja}
\affil{LESIA, Observatoire de Paris, Universit\'e PSL, CNRS, Sorbonne
Universit\'e, Universit\'e de Paris, 5 place Jules Janssen, 92195 Meudon,
France.}
\affiliation{Indian Institute of Astrophysics, 2nd Block, Koramangala, Bangalore - 560 034, India.}

\author{Prasad Subramanian}
\affiliation{Indian Institute of Science Education and Research, Pashan, Pune - 411 008, India.} % Institution(s)

\author{Madhusudan Ingale}

\affiliation{Indian Institute of Science Education and Research, Pashan, Pune - 411 008, India.}

\author{R. Ramesh}
\affiliation{Indian Institute of Astrophysics, 2nd Block, Koramangala, Bangalore - 560 034, India.} % Institution(s)

\author{Milan Maksimovic}
\affiliation{LESIA, Observatoire de Paris, Universit\'e PSL, CNRS, Sorbonne
Universit\'e, Universit\'e de Paris, 5 place Jules Janssen, 92195 Meudon,
France.}




\begin{abstract}

Various remote sensing observations have been used so far to probe the turbulent properties of the solar wind. Using the recently reported density modulation indices that are derived using angular broadening observations of Crab Nebula during 1952 - 2013, we measured the solar wind proton heating using the kinetic $\rm Alfv\acute{e}n$ wave dispersion equation. The estimated heating rates vary from $\approx 1.58 \times 10^{-14}$ to  $1.01 \times 10^{-8} ~\rm erg~ cm^{-3}~ s^{-1}$ in the heliocentric distance range 5 - 45 $\rm R_{\odot}$. Further, we found that heating rates vary with the solar cycle in correlation with density modulation indices. The models derived using in-situ measurements (for example, electron/proton density, temperature, and magnetic field) that the recently launched Parker Solar Probe observes (planned closest perihelia $\rm 9.86~ R_{\odot}$ from the center of the Sun) are useful in the estimation of the turbulent heating rate precisely. Further, we compared our heating rate estimates with the one derived using previously reported remote sensing and in-situ observations. \\

\end{abstract}


\section{Introduction}\label{sec:intro}

Even after decades of intense research, we do not know the precise solar wind heating mechanism and its acceleration. The in-situ observations confirm that solar wind undergoes extended non-adiabatic heating \citep{Freeman1988,Gazis1994, Matthaeus1999,Richardson2003}. More details on turbulent heating and solar wind acceleration are given by \citet{Cra2007, Cha2009b,Ver2010, Cra2013, Woo2014, Zan2018}. Interaction between counter-propagating \Alfven waves causes the wave energy to cascade into small scales. When the scale sizes which are perpendicular to the background magnetic field direction ($\lambda_{\perp}$) are comparable to the proton gyroradius ($\rho_p$), the wave energy begin to dissipate and thereby heats the plasma. It is known that when $\lambda_{\perp} \gg l_i = {v_A / \Omega_p}$, (where $l_i$ is the proton inertial length, $v_A$ is the \Alfven speed and $\Omega_p$ is the proton cyclotron frequency) the \Alfven waves are non-compressive. But there is a considerable evidence that when the scale sizes are in the range $\rho_p \lesssim \lambda_{\perp} \lesssim l_i$, they are compressive \citep[e.g.,][]{Harmon1989, Hollweg1999, Cha2009}. Also, the waves become dispersive and damp for the scales $\lambda_{\perp} \lesssim \rho_p$. 

In order to estimate the heating rates in the solar wind, we used the recently reported density modulation indices ($\epsilon_N = {\delta_N / N}$, where $\delta_N$ is the rms density fluctuations and `N' is the ambient background density) derived using the Crab Nebula occultation observations carried out during 1952 - 2013 \citep{Mac1952, Sle1959, Hew1963, Eri1964, Ble1972, Den1972, Sas1974, Arm1990, Ana1994, Subramanian2000, Ram2001, Sas2016, Sas2017, Sas2019a}. In this technique, when a radio point source (in this study Crab Nebula), observed through the foreground solar wind (in June of every year), we can have the following observations: (i) the radio source’s angular broadening increases due to the turbulent medium’s scattering, (ii) since we observe the radio sources whose flux density is constant over a long time, the peak flux density decreases as the source size increases; but the integrated flux density remains constant, (iii) the radio sources broaden anisotropically for the heliocentric distance below 10 $\rm R_{\odot}$, and thus we can measure the parameter anisotropy (i.e., the ratio between the major to the minor axis of radio source) \citep{Ble1972, Den1972, Sas2017}, and (iv) position angle of the major axis of the source (measured from the north through the east). Having such observations, \citet{Sas2016} have derived the density modulation indices in the heliocentric distance 5 - 45 $\rm R_{\odot}$. In this article, we use those density modulation indices to measure the proton heating rate ($\epsilon_{k_i}$) by making use of kinetic \Alfven wave dispersion equations (see \S \ref{sec:hr}) and compared the results with the recent reports that are measured using angular broadening \citep{Sas2017} and interplanetary scintillation observations \citep{Bis14,Ing2015b}. We also compare our results with the recently reported heating rates derived using the in-situ observations of \citet{Adh2020} and \citet{Ban2020}. Further, we report the way heating rates vary with the heliocentric distance and the way it vary with the solar cycle. 

\section{Observations}
\begin{figure}[!ht]
\centerline{\includegraphics[width=17cm]{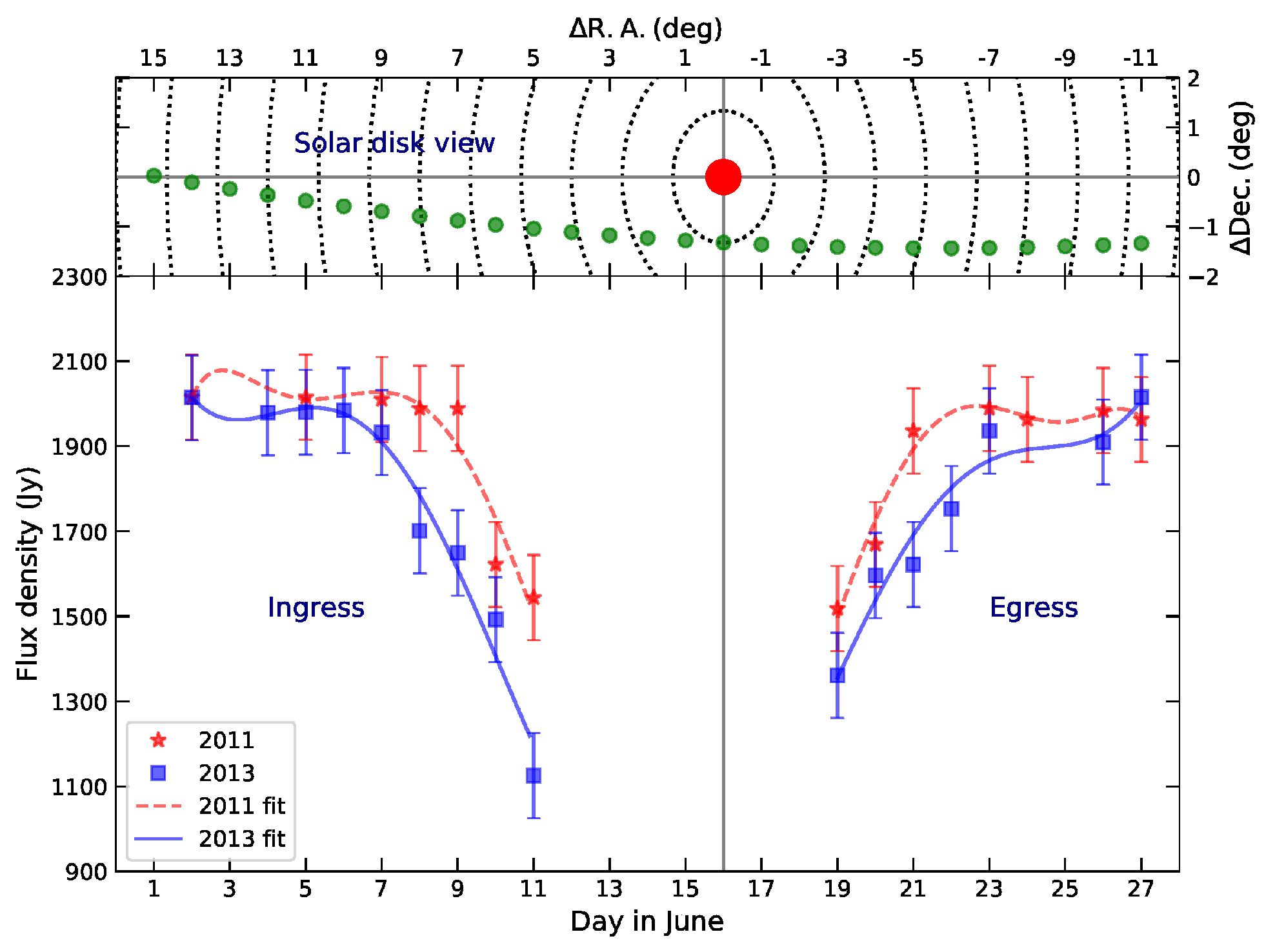}}
\caption{(top) Solar disk view of the Crab Nebula occultation. The red circle indicates 
the Sun and green circles represent the location of Crab Nebula with respect to the Sun in different days of June. $\Delta R.A.$ and $\Delta Dec.$ are the offset distances of Crab Nebula from the Sun in right ascension and declination, respectively. The closest dotted concentric circle around the Sun has a radius of $5~R_{\odot}$ and the radii of the rest of the circles differ from their adjacent ones by $5~R_{\odot}$. (bottom) The observed flux densities of the Crab Nebula on different days during its occultation by the solar corona. The periods before and after June 16th correspond to the ingress and egress, respectively. The markers `*' and `square' indicate the observations carried out in June 2011 and June 2013, respectively. The minimum detectable flux density of the GRAPH $\approx 100$ Jy is used as the error associated with these measurements.}
\label{fig:dv}
\end{figure}

\begin{figure*}[!ht]
\centerline{\includegraphics[width=18cm]{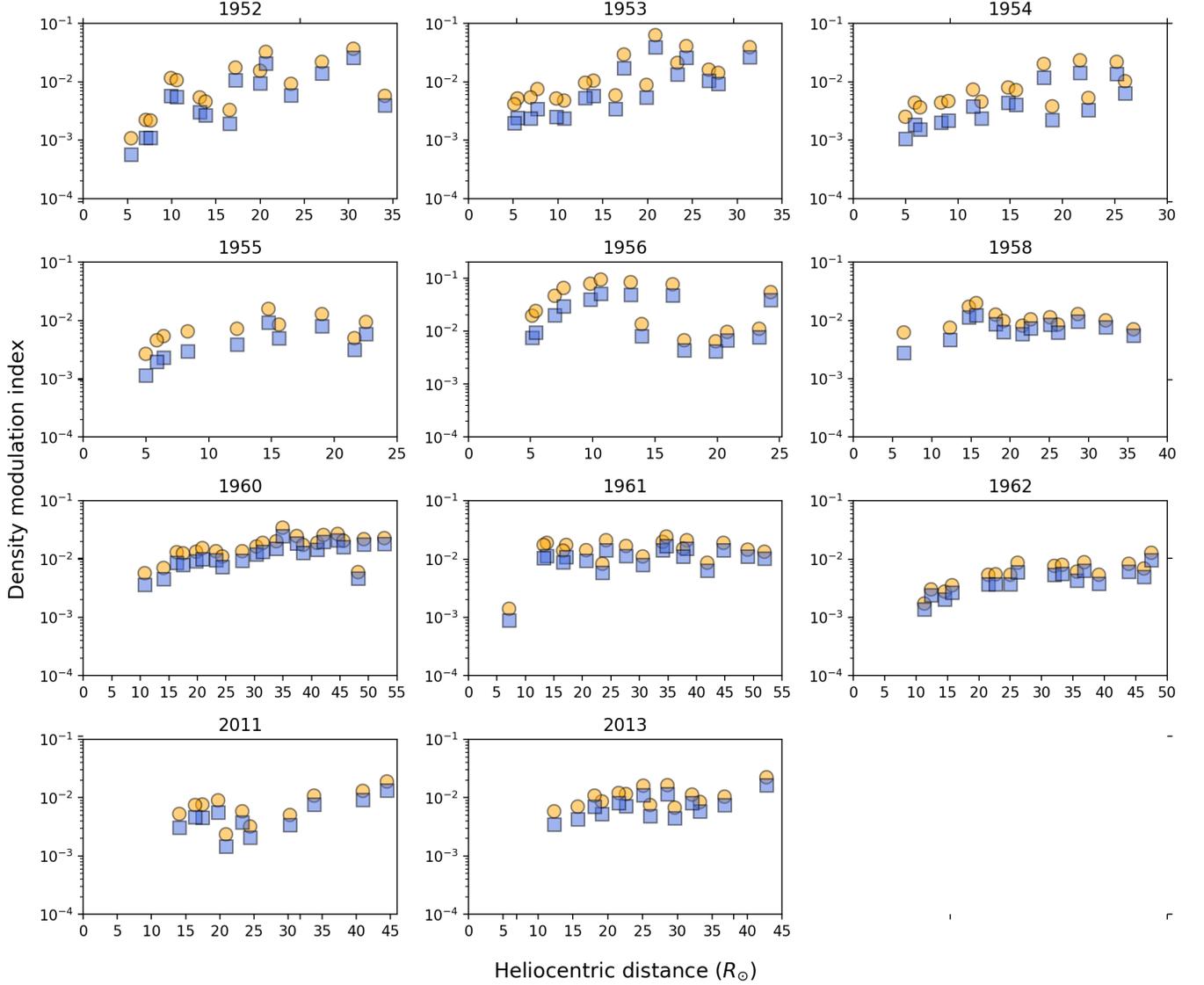}}
\caption{Heliocentric dependence of the density modulation index in different years. The markers `circle' and `square' indicate modulation indices that are derived using the proton inertial length and proton gyroradius model, respectively. %The marker `*' shows the empirically derived proton temperatures at a given heliocentric distance.
}
\label{fig:mi}
\end{figure*}

The angular broadening of the Crab Nebula is first observed by \citet{Mac1952}. Since then, many authors have reported similar observations as previously mentioned (see \S \ref{sec:intro}). In this article, we present results derived using data obtained by the Gauribidanur radioheliograph (GRAPH) during 2011 - 2013 \citep{Ramesh1998, Ram2011, Ramesh2014} and other historical observations carried out during 1952 - 1963 \citep{Mac1952, Hew1957, Hew1958, Hew1963, Sas2016}. For instance, Figure \ref{fig:dv} shows the observation of GRAPH carried out at 80 MHz over an interferometer baseline of 1600 meters. The top panel shows the schematic of the Crab Nebula occultation technique. The red and green circles indicate the sun and location of the Crab Nebula on different days of June in 2011 and 2013. The bottom panel shows the decrement in flux density as the Crab Nebula ingresses and becomes invisible during 12 - 18 June and then increments as it egresses. The flux density during 2013 is lower (compared to 2011) as it corresponds to the solar maximum. We make a note that the latter observations are carried out over interferometer baselines in the range 60 - 1000 meters and the frequency range 26-158 MHz. Therefore, \citet{Sas2016} have scaled these structure functions to the largest baseline of GRAPH ($1600$ meters; before the extension) and its routinely observed frequency 80 MHz using the general structure-function (see \S \ref{sec:dmi}). For the sake of completeness, we summarize a method using which \citet{Sas2016} have derived the density modulation indices (see Figure \ref{fig:mi}) and the way we have measured proton heating rate in the following sections. 

\section{Results and Discussions}\label{sec:dmi}

In the solar wind, turbulent density inhomogeneities play a vital role in scattering of the radio waves \citep{Col1989, Yam98, Bis14, Mug2017, Kru2018, Sas2019b, Kru2020}. Such inhomogeneities are represented by a spatial power spectrum. It comprises a power-law together with an exponential turnover at the inner scale. 
In the case of isotropic medium, the turbulent spatial power spectrum ($P_{\delta N}$) is defined as \citep{Bas94, Ing2015a}, 

\begin{equation}\label{eq:ss}
P_{\delta N}(k, R) = C_{N}^{2}(R) k^{-\alpha} \times \exp{-(kl_{i}(R) / 2 \pi)}^{2},
\end{equation}

where $k$ is the wavenumber, $l_i$ is the inner/ dissipation scale, and $C_{N}^{2}$ is the amplitude of density turbulence. It is worth mentioning that the injected large-scale energy in the solar wind breaks up into smaller scales until it is dissipated by heating the protons via gyro-resonant interactions. Also, we make a note that the scales at which the energy is injected are called `outer scales', and the scales at which the dissipation happens are called `inner scales' \citep{Kul2005}. Using remote sensing observations, it is found that, at large scales, the density spectrum follows the Kolmogorov scaling law with $\alpha=11/3$ \citep{Col1989, Spa02}. However, at small scales, the spectrum flattens to $\alpha=3$ \citep{Col1989}. In this article, since we are interested in the density fluctuations and proton heating rate near the dissipation scales, we have used $\alpha = 3$.
We make a note here that $C_N^2$ are measured for both proton inertial scale model  \citep{Col1989,Lea1999,Lea2000,Smi2001,Che2014,Bru2014,Sas2019a} and proton gyroradius model \citep{Bal2005, Sah13, Bis14, Che14, Sas2019a}. Note that \citet{Sas2016} measured the $C_N^2$ for two cases of proton temperatures $T_i = 10^5$ K and $T_i = 10^6$ K.

\subsection{Measurement of phase structure function}

A plane wave from a distant radio point source observed through the solar wind experiences loss of spatial and temporal coherence due to the refraction and scattering caused by the density inhomogeneities. The spatial coherence of the plane wave observed through the scattering medium (i.e., solar wind) is described by the mutual coherence function ($\Gamma(s)$), which is in turn related to the phase structure function ($D_{\phi}(s)$). We make a note that $D_{\phi}(s)$ provides the information to the extent to which ideal point source is broadened and 
it contains information about the spectrum of density turbulence. 
%A coherent wave observed through the turbulence medium experiences loss of spatial and temporal coherence due to the refraction and scattering caused by the density inhomogeneities associated the turbulence \citep{Ing2015}. 
In general, the phase structure function is defined as \citep{Col1989, Bas94, Ing2015a}, 

\begin{equation}
D_{\phi}(s) = \langle [\phi(r) - \phi (r+s)]^2 \rangle,
\end{equation}

where, $\langle~\rangle$ indicates the time average, `s' is the baseline of an interferometer, and $\phi(r)$ and $\phi(r+s)$ are the geometric phase delays in the line-of-sight direction through a turbulent medium at positions $r$ and $r+s$.

Using the Crab Nebula occultation observations we measure ($\Gamma(s)$) using, 

\begin{equation}
\Gamma(s) = {V(s) \over V(0)},
\end{equation}

where, V(s) is the peak flux density of the Crab Nebula observed through the scattering medium over a baseline `s', and V(0) is the flux density over a ``zero-length'' baseline. The quantity V(0) is measured when the Crab Nebula is far from the solar disk and is unresolved; $V(0) \approx 2015$ Jy at 80 MHz \citep{Bra70a, Mcl1985, Sas2017}.

By knowing the $\Gamma(s)$, we measured the density structure function ($D_{\phi}(s)$) using \citep{Pro1975,Ish1978,Col1989,Arm1990}, 

\begin{equation}
D_{\phi}(s)=-2 ln \Gamma(s)=-2ln\left[V(s)/V(0)\right].
\label{eq:struct}
\end{equation}

\subsection{The amplitude of density turbulence spectrum ($C_N^2$)}\label{sec:cn2}

By knowing the structure functions, we measured the amplitude of the turbulence ($C_N^2$) using the General Structure Function (GSF) \citep{Ing2015a, Sas2016, Sas2017}.
The GSF is defined as follows, 

{\begin{eqnarray}
\label{eq:gsf}
\nonumber
{D_\phi(s)} = \frac{8 \pi^2 r_e^2 \lambda^2 \Delta L}{\rho~ 2^{\alpha-2}(\alpha-2)} {\Gamma \bigg( 1 - {{\alpha-2} \over 2} \bigg)}
	    {{C_N^2 (R) l_i^{\alpha-2}(R)} \over {(1 - f_p^2 (R) / f^2)}} \\
	     {\times \bigg\{ { _1F_1} {\bigg[ - {{\alpha-2} \over 2},~1,~ - \bigg( {s \over l_i(R)} \bigg)^2 \bigg]} -1 \bigg\}} \, \, \, \, {\rm rad}^{2},
\end{eqnarray}}

where ${ _1F_1}$ is the confluent hyper-geometric function, $r_e$ is the classical electron radius, $\lambda$ is the observing wavelength, $R$ is the heliocentric distance (in units of $R_{\odot}$), $\Delta L$ is the thickness of the scattering medium ($\approx (\pi/2) R_{0}$, where $R_0$ is the impact parameter related to the projected heliocentric distance of the Crab Nebula), $f_p$ and f are the plasma and observing frequencies, respectively and the quantity $l_i$ is the inner scale. 

In order to evaluate the inner scales we used the following two prescriptions that are widely used in the literature. The first prescription envisages proton cyclotron damping by \Alfven waves. The inner scales measured using this mechanism are called proton inertial lengths \citep{Col1989,Lea1999,Lea2000,Smi2001,Che2014,Bru2014,Sas2019a} which can be written as,

\begin{equation}\label{eq:inner}
l_i(R) = v_A(R) / \Omega_p(R) = 2\pi/k_i(R)=228 / \sqrt{N_{e}(R)}\, \, \, {\rm km},
\end{equation}
 
where $N_{e}$ is the electron density in ${\rm cm}^{-3}$, $k_i$ is the wavenumber, $v_A$ is the $\rm Alfv\acute{e}n$ speed and $\Omega_p$ is the proton gyrofrequency.

The electron density ($N_e$) is estimated using the Leblanc density model \citep{Leb1998}:
\begin{equation}
N_e(R) = 7.2~R^{-2} + 1.95 \times 10^{-3}~R^{-4} + 8.1\times 10^{-7}~R^{-6} \,\,\,\, {\rm cm}^{-3}.
\label{leblanc}
\end{equation}

where `R' is the heliocentric distance in units of astronomical units (AU, 1 AU = $215 R_{\odot}$). 

In the second prescription, the inner scales are measured assuming proton gyroradius model in which dissipation is expected to happen at scales comparable to the proton gyroradius, $\rho_i=V_p / \Omega_p$, where $V_p$ is the proton speed and $\Omega_p$ is the proton gyrofrequency \citep{Gol2015}. The proton gyroradius scales are measured using \citep{Bal2005, Sah13, Bis14, Che14, Sas2019a}:

\begin{equation}
\rho_i(R) = 1.02 \times 10^2 \mu^{1/2} T_i^{1/2} B(R)^{-1} \,\,\,\, {\rm cm},
\label{ion_gyro}
\end{equation}

where $\mu (\equiv m_i/m_p)$ is the ion mass (in units of the proton mass), 
$T_i$ is the proton temperature (in eV) derived using following relations \citep{Ven2018},

\begin{equation}\label{eq:tp1}
T_{med}(SSN, R) = (197 \times SSN + 57300) \times R^{-1.10} ~~K
\end{equation}

\begin{equation}\label{eq:tp2}
T_{avg}(SSN, R)=1.654 \times T_{med}(SSN, R),
\end{equation}

where, $T_{med}$ and $T_{avg}$ are the median and average proton temperatures in K, and SSN is the sunspot number. We make a note that in this article, we have used the revised sunspot  number\footnote{\url{http://www.sidc.be/silso/datafiles}} \citep{Cle2016}.

Interplanetary magnetic field (B in Gauss) is measured using the Parker spiral magnetic field model in the ecliptic plane \citep{Wil1995},

\begin{equation}
B(R) = 3.4 \times 10^{-5}R^{-2}(1+R^2)^{1/2} \,\,\, {\rm Gauss}.
\label{imf}
\end{equation}

\subsection{Estimating the density modulation index ($\epsilon_{N_e}=\delta N_{k_i} / N_e$)}\label{lab:densmod}
The density fluctuations $\delta N_{k_i}$ at the inner scale and spatial power spectrum (Equation \ref{eq:ss}) are related as follows \citep{Cha2009}

\begin{equation}\label{eq:deltn}
{\delta}N_{k_i}^2(R) \sim 4 \pi k_i^3 P_{\delta N} (R, k_i) = 4 \pi C_{N}^{2}(R) k_i^{3 - \alpha} e^{-1} \,,
\end{equation}

where $k_{i} \equiv 2 \pi/l_{i}$. 

By knowing the ${\delta}N_{k_i}$ and the background electron density 
($N_{e}$, \S~3.1), the density modulation index ($\epsilon_{N_e}$) can be measured using, 

\begin{equation}\label{eq:df}
\epsilon_{N_e}(R) \equiv {~\delta N_{k_{i}}(R) \over N_{e}(R)}. 
\end{equation}
 
For the sake of completeness, the measured density modulation indices and its variation with heliocentric distance is shown in Figure \ref{fig:mi}  \citep{Sas2016}. Similarly, solar cycle dependence of the density modulation indices is shown in upper panel of Figure \ref{fig:sc} \citep{Sas2016}. Further, assuming the kinetic \Alfven wave dispersion equation we derived the heating rate. 

\subsection{Solar wind heating rate}\label{sec:hr}

\begin{figure*}[!ht]
\centerline{\includegraphics[width=18cm]{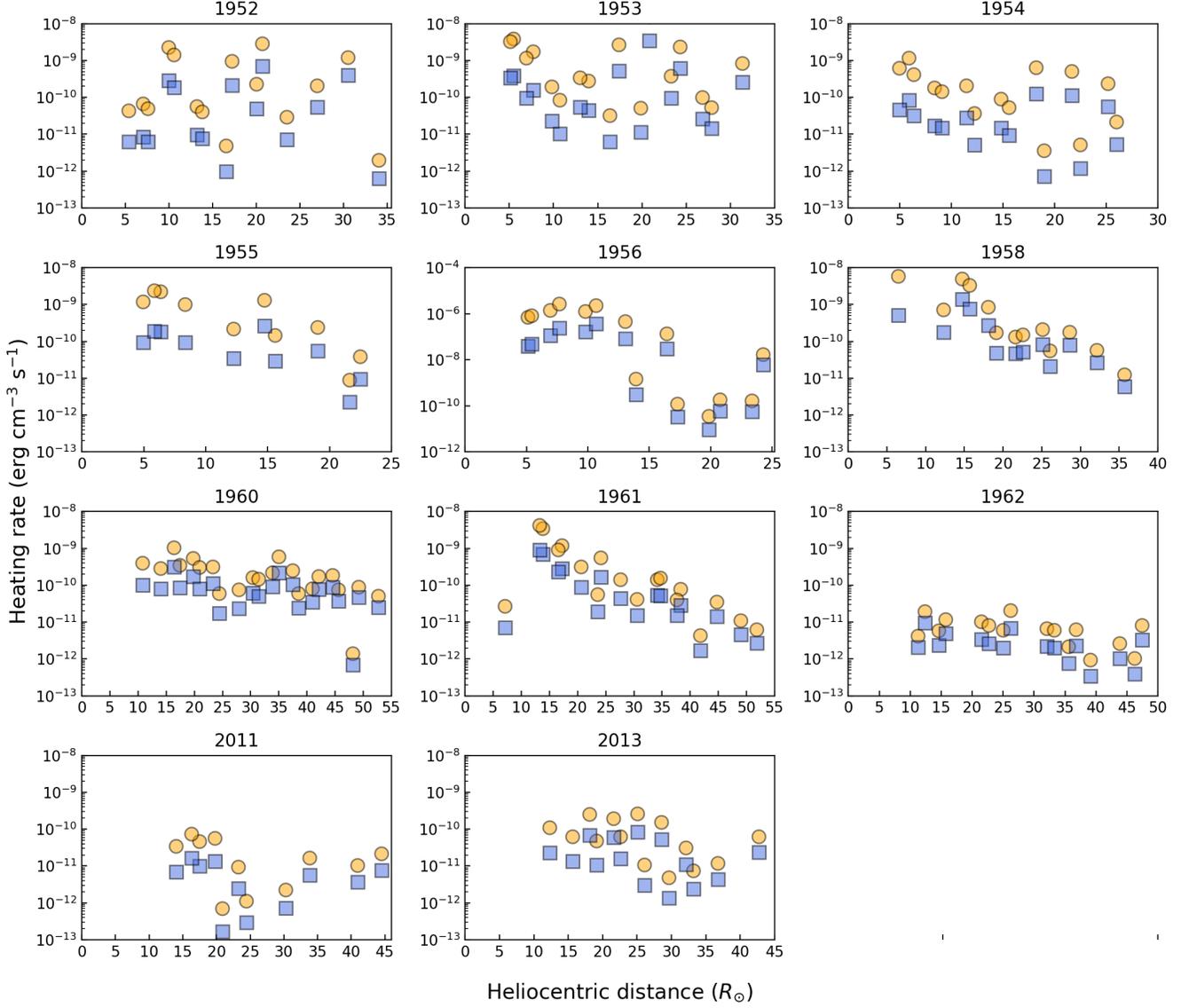}}
\caption{Heliocentric dependence of the proton heating rate in different years. The markers `circle' and `square' indicate proton heating rates that are derived using the proton inertial length and proton gyroradius model, respectively. %The marker `*' shows the empirically derived proton temperatures at a given heliocentric distance.
}
\label{fig:helio}
\end{figure*}

In this paper, we used the density modulation indices  ($\epsilon_{N_e}$) derived using the above method (see \S \ref{lab:densmod}) to measure the heating rates. Following \citet{Cha2009, Sas2017}, we assume that density fluctuations at small scales are manifestations of low frequency, 
oblique ($k_{\perp} \gg k_{\parallel}$), $\rm Alfv\acute{e}n$ wave turbulence and are often referred to 
kinetic $\rm Alfv\acute{e}n$ waves. 
Here, the quantities $k_{\perp}$ 
and $k_{\parallel}$ are the components of the wave vector k
in perpendicular and parallel direction to the background large-scale magnetic field, respectively.

As previously discussed we envisage a situation where the ``balanced'' counter propagating $\rm Alfv\acute{e}n$ 
waves (i.e. with zero helicity) cascade and resonantly damps on the 
protons at the inner scale and thereby heats the solar wind. 
Because of the passive mixing of the $\rm Alfv\acute{e}n$ waves with other modes at the inner scale
our proton heating rate measurements provide an upper limit. The proton heating rate (i.e. the turbulent energy cascade rate) at inner scales is \citep{Hollweg1999, Cha2009, Ing2015b}, 

\begin{equation}\label{eq:hr}
\epsilon_{k_i}(R)=c_0 \rho_p k_i(R) \delta v_{k_i}^3(R) ~ \rm erg ~cm^{-3}~s^{-1} \, ,
\end{equation}

where, $\rho_p=m_pN_e(R)~\rm g~ cm^{-3}$ with $m_p$ is the proton mass [in grams],  $k_i=2 \pi/l_i$ and $\delta v_{k_i}$ are the wavenumber and  magnitude of turbulent velocity fluctuations at inner scales, respectively. The dimensional less quantity $c_0$ is assumed to be 0.25 \citep{How2008, Cha2009, Sas2017}. 

By knowing the $\epsilon_{N_e}$, we calculated $\delta v_{k_i}$ using the 
kinetic $\rm Alfv\acute{e}n$ wave dispersion relation \citep{How2008,Cha2009,Ing2015b, Sas2017}

\begin{eqnarray}\label{eq:rmsv}
 \delta v_{k_i}(R)=\Bigg({1+{\gamma_i k_i^2(R) \rho_i^2(R)} \over {k_i(R) l_i(R)}} \Bigg) \epsilon_{N_e} (R, k_i) v_A(R) \, .
\end{eqnarray}

where, the adiabatic index $\gamma_i$ is taken to be 1 \citep{Cha2009, Sas2017}. 

The $\rm Alfv\acute{e}n$ speed ($v_A$) in the solar wind is measured using, 

\begin{equation}\label{eq:va}
v_A(R)=2.18\times 10^{11} \mu^{-1/2} N_e^{-1/2}(R)B(R) ~\rm cm~s^{-1}, 
\end{equation}

The magnetic field strength (B) is estimated using the Parker spiral magnetic field in the ecliptic 
plane using \citep{Wil1995}, 

\begin{equation}\label{eq:parker}
B(R)= 3.4 \times 10^{-5} R^{-2} (1+R^2)^{1/2} ~ \rm Gauss, 
\end{equation}

where, `R' is the heliocentric distance in units of AU. 

\begin{figure}[!ht]
\centerline{\includegraphics[width=15cm]{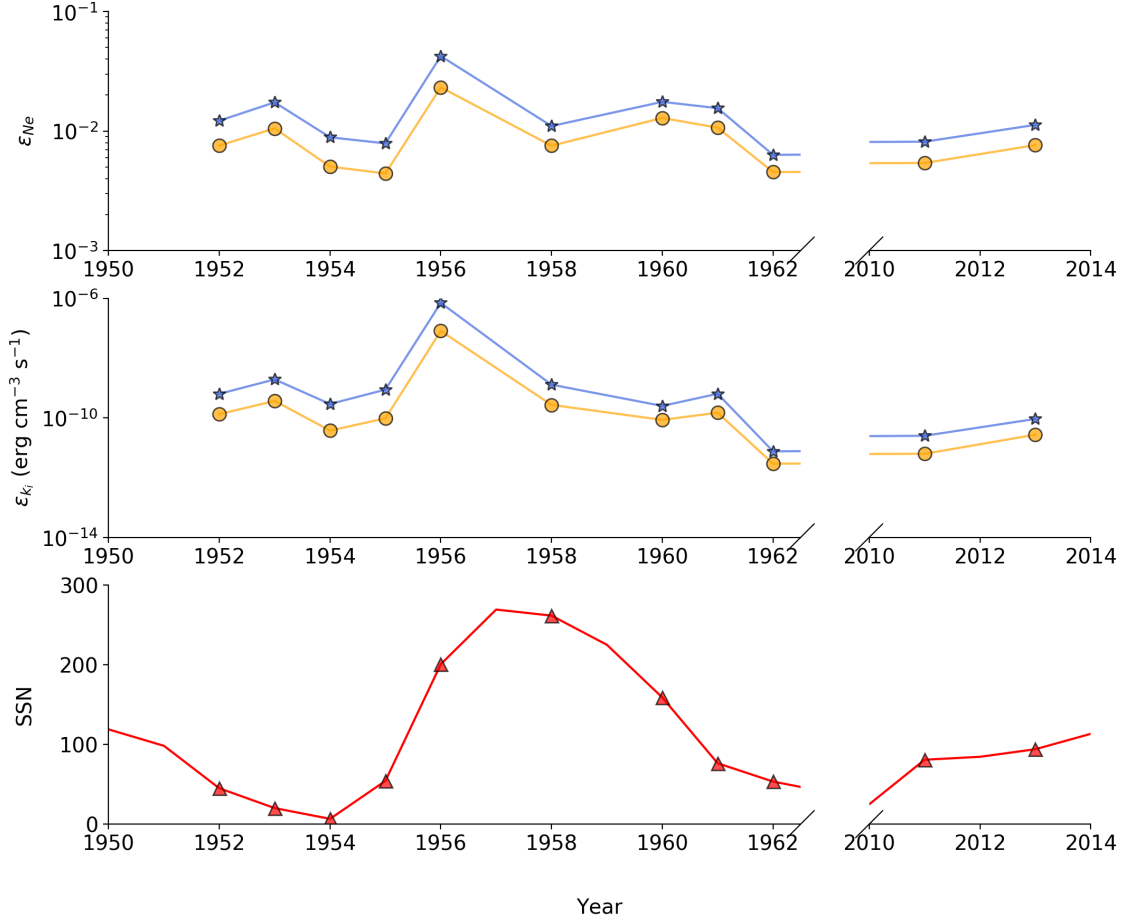}}
\caption{(top) Solar cycle dependence of the modulation index averaged over the heliocentric distance $5-45~ R_{\odot}$.
(middle) Solar cycle dependence of the proton heating rate averaged over the heliocentric distance $5-45~ R_{\odot}$. The measurements for various inner scale models, i.e., proton inertial length and proton gyroradius models are indicated with the symbols `circle' and `square', respectively. (bottom) The solid line shows the yearly averaged sunspot number \citep{Cle2016} and the symbol `triangle' indicates the sunspot number in which we have the radio observations. The Figure shows a clear solar cycle dependence of the proton heating rate.}
\label{fig:sc}
\end{figure}

The derived proton heating rates in different years are shown in Figure \ref{fig:helio} and we found that heating rates vary from $\approx 1.58 \times 10^{-14}$ to  $1.01 \times 10^{-8} ~\rm erg~ cm^{-3}~ s^{-1}$ over the heliocentric distances 5 - 45 $\rm R_{\odot}$. The markers `circle' and `square' indicate proton heating rates derived assuming different inner scale models - proton inertial length and proton gyroradius model. For latter case, the inner scales are measured using the proton temperature derived using equations \ref{eq:tp1} and \ref{eq:tp2}. %Furthermore, we have plotted these proton temperatures (indicated by markers `*') in different years as shown in Figure \ref{fig:helio} and found that the proton heating rate correlates with the proton temperature.

At 5 $\rm R_{\odot}$, in the coronal holes (i.e., in the fast solar wind), the estimated proton heating rates range from $2 \times 10^{-10}$ and $1.4 \times 10^{-8} ~\rm~ erg~ cm^{-3} ~s^{-1}$ \citep{Cha2009}. Similarly, at 1 AU the estimated heating rate is $5 \times 10^{-16} \rm~ erg ~cm^{-3}~ s^{-1}$ \citep{Cha2009}. The heating rates derived assuming density fluctuations are due to the kinetic \Alfven waves in the heliocentric distance range 2-174 $\rm R_{\odot}$ using interplanetary scintillation (IPS) observations \citep{Hew1963,Man2000,Jan11,Sas2019b} are $3 \times 10^{-8} ~\rm erg~ cm^{-3}~ s^{-1}$ (during solar maximum) and $\approx 10^{-15} ~\rm erg~ cm^{-3}~ s^{-1}$ (during solar minimum) consistent with our estimates \citep{Ing2015b}. Using two-dimensional imaging angular broadening observations of Crab Nebula, the measured heating rates are varied from $2.2 \times 10^{-13}$ to $ 1.0 \times 10^{-11} ~\rm erg~ cm^{-3}~ s^{-1}$ in the projected heliocentric distance range $9 - 20~ R_{\odot}$ \citep{Sas2017}. The recently reported heating rates in the heliocentric distance range 1.5 - 4.0 $R_{\odot}$ varied from $\approx 3.31 \times 10^{-10}$ to $4.5 \times 10^{-7}\rm~ erg~ cm^{-3}~ s^{-1}$ \citep{Ste2020}. Further, the author extrapolated these heating rates to the distances 0.3-0.6 AU and it range from $\approx 10^{-15}$ to $10^{-14}\rm~ erg~ cm^{-3}~ s^{-1}$ and at 1 AU, the extrapolated heating rates are few times of $10^{-16}\rm~ erg~ cm^{-3}~ s^{-1}$.

Using in-situ measurements by Parker Solar Probe, \citet{Ban2020} estimated energy transfer rates of $8.7 \pm 0.3 \times 10^{-13}~\rm erg~ cm^{-3}~ s^{-1}$ at 36 $R_{\odot}$ and $5.8 \pm 1.3 \times 10^{-14} ~\rm erg~ cm^{-3}~ s^{-1}$ at 54 $R_{\odot}$. They originally have quoted numbers in units of ${\rm J\,kg^{-1}\,s^{-1}}$. We have multiplied their numbers by $ N_{p}\, m_{p}$ (where $N_{p}$ is the solar wind density derived using Leblanc model \citep{Leb1998} and $m_{p}$ is the proton mass) to arrive at heating rates in units of ${\rm erg\,cm^{-3}\,s^{-1}}$. By comparison, the proton heating rate at 36 $R_{\odot}$ from our results (see Figure \ref{fig:helio}) range from $\approx 2.8 \times 10^{-10}$ to $\approx 7.4 \times 10^{-13}$ ${\rm erg\,cm^{-3}\,s^{-1}}$. 
Similarly, \citet{Adh2020} reported that heating rates due to quasi-2D turbulence in the heliocentric distance $\approx 1.6 - 100 ~R_{\odot}$ range from $1.06 \times 10^{-4}~\rm~to~ 1.73 \times 10^{-14}~ erg~ cm^{-3}~ s^{-1}$. Authors also reported that the heating rate due to the nearly in-compressible/slab turbulence in the heliocentric distance $\approx 1.3 - 100 ~R_{\odot}$ range from $4.24 \times 10^{-7}~\rm~to~ 1.11 \times 10^{-14}~ erg~ cm^{-3}~ s^{-1}$. A summary of these proton heating rates is given in Table \ref{tab:hr123}.

\begin{table}[h!]
\centering
 \begin{tabular}{c c c c}
 	\hline
 	S.No & R             & Proton heating rate                             & References       \\
 	     & ($R_{\odot}$) & ($\rm erg ~cm^{-3}~ s^{-1}$)                    &                  \\ \hline
 	                           \multicolumn{4}{c}{Remote sensing}                             \\ \hline
 	 1   & 5 - 45        & $1.58 \times 10^{-14}$ -  $1.01 \times 10^{-8}$ & Present work     \\
 	 2   & 5             & $2 \times 10^{-10}$ - $1.4 \times 10^{-8}$      & \citet{Cha2009}  \\
 	 3   & 215           & $5 \times 10^{-16}$                             & \citet{Cha2009}  \\
 	 4   & 2  -174       & $3 \times 10^{-8}$ - $10^{-15}$                 & \citet{Ing2015b} \\
 	 5   & 9 - 20        & $2.2 \times 10^{-13}$ - $ 1.0 \times 10^{-11}$  & \citet{Sas2017}  \\
 	 6   & 1.5 - 4.0     & $3.31 \times 10^{-10}$ - $4.5 \times 10^{-7}$   & \citet{Ste2020}  \\
 	 7   & 64.5 - 129    & $10^{-15}$ - $10^{-14}$                         & \citet{Ste2020}  \\
 	 8   & 215           & $10^{-16}$                                      & \citet{Ste2020}  \\ \hline
 	                               \multicolumn{4}{c}{In-situ}                                \\ \hline
 	 9   & 36            & $8.7 \pm 0.3 \times 10^{-13}$                   & \citet{Ban2020}  \\
 	 10  & 54            & $5.8 \pm 1.3 \times 10^{-14} $                  & \citet{Ban2020}  \\
 	 11  & 1.6 - 100     & $1.06 \times 10^{-4}$ - $1.73 \times 10^{-14}$  & \citet{Adh2020}  \\
 	 12  & 1.3 - 100     & $4.24 \times 10^{-7}$ - $1.11 \times 10^{-14}$  & \citet{Adh2020}  \\ \hline
 \end{tabular}
\caption{Summary of proton heating rates in the solar wind}
\label{tab:hr123}
\end{table}

As the density modulation indices (see Figure \ref{fig:mi}) and heating rates (see Figure \ref{fig:helio}) are weakly dependent with heliocentric distance, we averaged the observations that are carried out in different years and plotted in Figure \ref{fig:sc}. The upper and middle panels of Figure \ref{fig:sc} are the averaged density modulation indices and proton heating rates for different inner scale models, respectively. The lower panel shows the yearly averaged sunspot number. Figure \ref{fig:sc} shows that the derived density modulation indices and heating rates closely follow the solar cycle. During solar maximum, the slow solar wind drives in all the directions and hence \citet{Sas2016} had justified the lower modulation index in 1958 (also refer to upper panel of Figure \ref{fig:sc}). Following the lower density modulation indices, heating rates are lower during the solar maximum.

\section{Summary and Conclusions}

In this article, we have used recently reported density modulation indices $\epsilon_N$ derived using angular broadening observations of Crab Nebula \citep{Sas2016}. The authors have studied the way $\epsilon_N$ vary with heliocentric distance and solar cycle (see Figure \ref{fig:mi} and \ref{fig:sc}). Using imaging observations of the Crab Nebula observed in 2016 and 2017, the proton heating rate in the solar corona at various heliocentric distances are reported \citep{Sas2017}. Using these $\epsilon_N$ values and the method discussed by \citet{Sas2017}, 
we measured the proton heating rate in different years. We found that during 1952 and 2013, the measured proton heating rate ranges from $\approx 1.58 \times 10^{-14}$ to  $1.01 \times 10^{-8} ~\rm erg~ cm^{-3}~ s^{-1}$
in the heliocentric distance $5-45~R_  {\odot}$ as shown in Figure \ref{fig:helio}. 
As the density modulation indices and heating rates weakly depend on the heliocentric distance, we averaged the year's entire observations.  
The upper and middle panels of Figure \ref{fig:sc} show the way the density modulation indices and proton heating rate vary in different years. The lower panel shows the yearly averaged sunspot number in the respective years. 
Hence we conclude that both density modulation indices and proton heating rate in the solar wind correlates with the solar cycle. The in-situ measurements and thus the derived models (for example, electron / proton density, temperature, and magnetic field) using the Parker Solar Probe which has already covered the heliocentric distance range of 25 $\rm R_{\odot}$ and planned to reach as close as 9.86 $\rm R_{\odot}$ \citep{Fox2016} plays a significant role in better understanding proton heating rates and thus the solar wind acceleration.

\section*{Acknowledgment}
KSR acknowledges the financial support from the Centre National d'\'{e}tudes Spatiales (CNES), France. KSR acknowledges O. Alexandrova for the useful discussions that helped in improving the manuscript. The sunspot number used in this article is credited to WDC-SILSO, Royal Observatory of Belgium, Brussels. We thank the referee for constructive suggestions and comments that helped in improving the manuscript. 
\bibliographystyle{aasjournal}
\bibliography{ms}

\begin{thebibliography}{}
\expandafter\ifx\csname natexlab\endcsname\relax\def\natexlab#1{#1}\fi

\bibitem[{{Adhikari} {et~al.}(2020){Adhikari}, {Zank}, \& {Zhao}}]{Adh2020}
{Adhikari}, L., {Zank}, G.~P., \& {Zhao}, L.~L. 2020, \apj, 901, 102

\bibitem[{{Anantharamaiah} {et~al.}(1994){Anantharamaiah}, {Gothoskar}, \&
  {Cornwell}}]{Ana1994}
{Anantharamaiah}, K.~R., {Gothoskar}, P., \& {Cornwell}, T.~J. 1994, Journal of
  Astrophysics and Astronomy, 15, 387

\bibitem[{{Armstrong} {et~al.}(1990){Armstrong}, {Coles}, {Rickett}, \&
  {Kojima}}]{Arm1990}
{Armstrong}, J.~W., {Coles}, W.~A., {Rickett}, B.~J., \& {Kojima}, M. 1990, The
  Astrophysical Journal, 358, 685

\bibitem[{{Bale} {et~al.}(2005){Bale}, {Kellogg}, {Mozer}, {Horbury}, \&
  {Reme}}]{Bal2005}
{Bale}, S.~D., {Kellogg}, P.~J., {Mozer}, F.~S., {Horbury}, T.~S., \& {Reme},
  H. 2005, \prl, 94, 215002

\bibitem[{{Bandyopadhyay} {et~al.}(2020){Bandyopadhyay}, {Goldstein}, {Maruca},
  {Matthaeus}, {Parashar}, {Ruffolo}, {Chhiber}, {Usmanov}, {Chasapis},
  {Qudsi}, {Bale}, {Bonnell}, {Dudok de Wit}, {Goetz}, {Harvey}, {MacDowall},
  {Malaspina}, {Pulupa}, {Kasper}, {Korreck}, {Case}, {Stevens}, {Whittlesey},
  {Larson}, {Livi}, {Klein}, {Velli}, \& {Raouafi}}]{Ban2020}
{Bandyopadhyay}, R., {Goldstein}, M.~L., {Maruca}, B.~A., {et~al.} 2020, \apjs,
  246, 48

\bibitem[{{Bastian}(1994)}]{Bas94}
{Bastian}, T.~S. 1994, The Astrophysical Journal, 426, 774

\bibitem[{{Bisoi} {et~al.}(2014){Bisoi}, {Janardhan}, {Ingale}, {Subramanian},
  {Ananthakrishnan}, {Tokumaru}, \& {Fujiki}}]{Bis14}
{Bisoi}, S.~K., {Janardhan}, P., {Ingale}, M., {et~al.} 2014, The Astrophysical
  Journal, 795, 69

\bibitem[{{Blesing} \& {Dennison}(1972)}]{Ble1972}
{Blesing}, R.~G., \& {Dennison}, P.~A. 1972, Proceedings of the Astronomical
  Society of Australia, 2, 84

\bibitem[{{Braude} {et~al.}(1970){Braude}, {Lebedeva}, {Megn}, {Ryabov}, \&
  {Zhouck}}]{Bra70a}
{Braude}, S.~Y., {Lebedeva}, O.~M., {Megn}, A.~V., {Ryabov}, B.~P., \&
  {Zhouck}, I.~N. 1970, Astrophysics Letters, 5, 129

\bibitem[{{Bruno} \& {Trenchi}(2014)}]{Bru2014}
{Bruno}, R., \& {Trenchi}, L. 2014, \apjl, 787, L24

\bibitem[{{Chandran} \& {Hollweg}(2009b)}]{Cha2009b}
{Chandran}, B. D.~G., \& {Hollweg}, J.~V. 2009b, \apj, 707, 1659

\bibitem[{{Chandran} {et~al.}(2009a){Chandran}, {Quataert}, {Howes}, {Xia}, \&
  {Pongkitiwanichakul}}]{Cha2009}
{Chandran}, B.~D.~G., {Quataert}, E., {Howes}, G.~G., {Xia}, Q., \&
  {Pongkitiwanichakul}, P. 2009a, The Astrophysical Journal, 707, 1668

\bibitem[{{Chen} {et~al.}(2014{\natexlab{a}}){Chen}, {Leung}, {Boldyrev},
  {Maruca}, \& {Bale}}]{Che2014}
{Chen}, C.~H.~K., {Leung}, L., {Boldyrev}, S., {Maruca}, B.~A., \& {Bale},
  S.~D. 2014{\natexlab{a}}, \grl, 41, 8081

\bibitem[{{Chen} {et~al.}(2014{\natexlab{b}}){Chen}, {Leung}, {Boldyrev},
  {Maruca}, \& {Bale}}]{Che14}
---. 2014{\natexlab{b}}, Geophysical Research Letters, 41, 8081

\bibitem[{{Clette} {et~al.}(2016){Clette}, {Lef{\`e}vre}, {Cagnotti},
  {Cortesi}, \& {Bulling}}]{Cle2016}
{Clette}, F., {Lef{\`e}vre}, L., {Cagnotti}, M., {Cortesi}, S., \& {Bulling},
  A. 2016, \solphys, 291, 2733

\bibitem[{{Coles} \& {Harmon}(1989)}]{Col1989}
{Coles}, W.~A., \& {Harmon}, J.~K. 1989, The Astrophysical Journal, 337, 1023

\bibitem[{{Cranmer}(2020)}]{Ste2020}
{Cranmer}, S.~R. 2020, arXiv e-prints, arXiv:2007.13180

\bibitem[{{Cranmer} {et~al.}(2007){Cranmer}, {van Ballegooijen}, \&
  {Edgar}}]{Cra2007}
{Cranmer}, S.~R., {van Ballegooijen}, A.~A., \& {Edgar}, R.~J. 2007, \apjs,
  171, 520

\bibitem[{{Cranmer} {et~al.}(2013){Cranmer}, {van Ballegooijen}, \&
  {Woolsey}}]{Cra2013}
{Cranmer}, S.~R., {van Ballegooijen}, A.~A., \& {Woolsey}, L.~N. 2013, \apj,
  767, 125

\bibitem[{{Dennison} \& {Blesing}(1972)}]{Den1972}
{Dennison}, P.~A., \& {Blesing}, R.~G. 1972, Proceedings of the Astronomical
  Society of Australia, 2, 86

\bibitem[{{Erickson}(1964)}]{Eri1964}
{Erickson}, W.~C. 1964, \apj, 139, 1290

\bibitem[{{Fox} {et~al.}(2016){Fox}, {Velli}, {Bale}, {Decker}, {Driesman},
  {Howard}, {Kasper}, {Kinnison}, {Kusterer}, {Lario}, {Lockwood}, {McComas},
  {Raouafi}, \& {Szabo}}]{Fox2016}
{Fox}, N.~J., {Velli}, M.~C., {Bale}, S.~D., {et~al.} 2016, \ssr, 204, 7

\bibitem[{{Freeman}(1988)}]{Freeman1988}
{Freeman}, J.~W. 1988, \grl, 15, 88

\bibitem[{{Gazis} {et~al.}(1994){Gazis}, {Barnes}, {Mihalov}, \&
  {Lazarus}}]{Gazis1994}
{Gazis}, P.~R., {Barnes}, A., {Mihalov}, J.~D., \& {Lazarus}, A.~J. 1994, \jgr,
  99, 6561

\bibitem[{{Goldstein} {et~al.}(2015){Goldstein}, {Wicks}, {Perri}, \&
  {Sahraoui}}]{Gol2015}
{Goldstein}, M.~L., {Wicks}, R.~T., {Perri}, S., \& {Sahraoui}, F. 2015,
  Philosophical Transactions of the Royal Society of London Series A, 373,
  20140147

\bibitem[{{Harmon}(1989)}]{Harmon1989}
{Harmon}, J.~K. 1989, \jgr, 94, 15399

\bibitem[{{Hewish}(1957)}]{Hew1957}
{Hewish}, A. 1957, The Observatory, 77, 151

\bibitem[{{Hewish}(1958)}]{Hew1958}
---. 1958, \mnras, 118, 534

\bibitem[{{Hewish} \& {Wyndham}(1963)}]{Hew1963}
{Hewish}, A., \& {Wyndham}, J.~D. 1963, \mnras, 126, 469

\bibitem[{{Hollweg}(1999)}]{Hollweg1999}
{Hollweg}, J.~V. 1999, \jgr, 104, 14811

\bibitem[{{Howes} {et~al.}(2008){Howes}, {Cowley}, {Dorland}, {Hammett},
  {Quataert}, \& {Schekochihin}}]{How2008}
{Howes}, G.~G., {Cowley}, S.~C., {Dorland}, W., {et~al.} 2008, Journal of
  Geophysical Research (Space Physics), 113, A05103

\bibitem[{{Ingale}(2015b)}]{Ing2015b}
{Ingale}, M. 2015b, arXiv e-prints, arXiv:1509.07652

\bibitem[{{Ingale} {et~al.}(2015a){Ingale}, {Subramanian}, \&
  {Cairns}}]{Ing2015a}
{Ingale}, M., {Subramanian}, P., \& {Cairns}, I. 2015a, \mnras, 447, 3486

\bibitem[{{Ishimaru}(1978)}]{Ish1978}
{Ishimaru}, A. 1978, {Wave propagation and scattering in random media. Volume 1
  - Single scattering and transport theory}, Vol.~1,
  doi:10.1016/B978-0-12-374701-3.X5001-7

\bibitem[{{Janardhan} {et~al.}(2011){Janardhan}, {Bisoi}, {Ananthakrishnan},
  {Tokumaru}, \& {Fujiki}}]{Jan11}
{Janardhan}, P., {Bisoi}, S.~K., {Ananthakrishnan}, S., {Tokumaru}, M., \&
  {Fujiki}, K. 2011, Geophysical Research Letters, 38, L20108

\bibitem[{{Krupar} {et~al.}(2018){Krupar}, {Maksimovic}, {Kontar}, {Zaslavsky},
  {Santolik}, {Soucek}, {Kruparova}, {Eastwood}, \& {Szabo}}]{Kru2018}
{Krupar}, V., {Maksimovic}, M., {Kontar}, E.~P., {et~al.} 2018, \apj, 857, 82

\bibitem[{{Krupar} {et~al.}(2020){Krupar}, {Szabo}, {Maksimovic}, {Kruparova},
  {Kontar}, {Balmaceda}, {Bonnin}, {Bale}, {Pulupa}, {Malaspina}, {Bonnell},
  {Harvey}, {Goetz}, {Dudok de Wit}, {MacDowall}, {Kasper}, {Case}, {Korreck},
  {Larson}, {Livi}, {Stevens}, {Whittlesey}, \& {Hegedus}}]{Kru2020}
{Krupar}, V., {Szabo}, A., {Maksimovic}, M., {et~al.} 2020, \apjs, 246, 57

\bibitem[{{Kulsrud}(2005)}]{Kul2005}
{Kulsrud}, R.~M. 2005, {Plasma physics for astrophysics}

\bibitem[{{Leamon} {et~al.}(2000){Leamon}, {Matthaeus}, {Smith}, {Zank},
  {Mullan}, \& {Oughton}}]{Lea2000}
{Leamon}, R.~J., {Matthaeus}, W.~H., {Smith}, C.~W., {et~al.} 2000, \apj, 537,
  1054

\bibitem[{{Leamon} {et~al.}(1999){Leamon}, {Smith}, {Ness}, \&
  {Wong}}]{Lea1999}
{Leamon}, R.~J., {Smith}, C.~W., {Ness}, N.~F., \& {Wong}, H.~K. 1999, \jgr,
  104, 22331

\bibitem[{{Leblanc} {et~al.}(1998){Leblanc}, {Dulk}, \& {Bougeret}}]{Leb1998}
{Leblanc}, Y., {Dulk}, G.~A., \& {Bougeret}, J.-L. 1998, \solphys, 183, 165

\bibitem[{{Machin} \& {Smith}(1952)}]{Mac1952}
{Machin}, K.~E., \& {Smith}, F.~G. 1952, Nature, 170, 319

\bibitem[{{Manoharan} {et~al.}(2000){Manoharan}, {Kojima}, {Gopalswamy},
  {Kondo}, \& {Smith}}]{Man2000}
{Manoharan}, P.~K., {Kojima}, M., {Gopalswamy}, N., {Kondo}, T., \& {Smith}, Z.
  2000, \apj, 530, 1061

\bibitem[{{Matthaeus} {et~al.}(1999){Matthaeus}, {Zank}, {Smith}, \&
  {Oughton}}]{Matthaeus1999}
{Matthaeus}, W.~H., {Zank}, G.~P., {Smith}, C.~W., \& {Oughton}, S. 1999, \prl,
  82, 3444

\bibitem[{{McLean} \& {Labrum}(1985)}]{Mcl1985}
{McLean}, D.~J., \& {Labrum}, N.~R. 1985, {Solar radiophysics: Studies of
  emission from the sun at metre wavelengths}

\bibitem[{{Mugundhan} {et~al.}(2017){Mugundhan}, {Hariharan}, \&
  {Ramesh}}]{Mug2017}
{Mugundhan}, V., {Hariharan}, K., \& {Ramesh}, R. 2017, Solar Physics, 292, 155

\bibitem[{{Prokhorov} {et~al.}(1975){Prokhorov}, {Bunkin}, {Gochelashvili}, \&
  {Shishov}}]{Pro1975}
{Prokhorov}, A.~M., {Bunkin}, F.~V., {Gochelashvili}, K.~S., \& {Shishov},
  V.~I. 1975, IEEE Proceedings, 63, 790

\bibitem[{{Ramesh}(2011)}]{Ram2011}
{Ramesh}, R. 2011, in Astronomical Society of India Conference Series, Vol.~2,
  Astronomical Society of India Conference Series

\bibitem[{{Ramesh}(2014)}]{Ramesh2014}
{Ramesh}, R. 2014, in Astron. Soc. India Conf. Ser., Vol.~13, Metrewavelength
  Sky, ed. J.~N. {Chengalur} \& Y.~{Gupta}, 19

\bibitem[{{Ramesh} {et~al.}(2001){Ramesh}, {Kathiravan}, \& {Sastry}}]{Ram2001}
{Ramesh}, R., {Kathiravan}, C., \& {Sastry}, C.~V. 2001, The Astrophysical
  Journal, Letters, 548, L229

\bibitem[{{Ramesh} {et~al.}(1998){Ramesh}, {Subramanian}, {Sundara Rajan}, \&
  Sastry}]{Ramesh1998}
{Ramesh}, R., {Subramanian}, K.~R., {Sundara Rajan}, M.~S., \& Sastry, C.~V.
  1998, Solar Phys., 181, 439

\bibitem[{{Richardson} \& {Smith}(2003)}]{Richardson2003}
{Richardson}, J.~D., \& {Smith}, C.~W. 2003, \grl, 30, 1206

\bibitem[{{Sahraoui} {et~al.}(2013){Sahraoui}, {Huang}, {Belmont}, {Goldstein},
  {R{\'e}tino}, {Robert}, \& {De Patoul}}]{Sah13}
{Sahraoui}, F., {Huang}, S.~Y., {Belmont}, G., {et~al.} 2013, The Astrophysical
  Journal, 777, 15

\bibitem[{{Sasikumar Raja} {et~al.}(2016){Sasikumar Raja}, {Ingale}, {Ramesh},
  {Subramanian}, {Manoharan}, \& {Janardhan}}]{Sas2016}
{Sasikumar Raja}, K., {Ingale}, M., {Ramesh}, R., {et~al.} 2016, Journal of
  Geophysical Research (Space Physics), 121, 11605

\bibitem[{{Sasikumar Raja} {et~al.}(2019b){Sasikumar Raja}, {Janardhan},
  {Bisoi}, {Ingale}, {Subramanian}, {Fujiki}, \& {Maksimovic}}]{Sas2019b}
{Sasikumar Raja}, K., {Janardhan}, P., {Bisoi}, S.~K., {et~al.} 2019b,
  \solphys, 294, 123

\bibitem[{{Sasikumar Raja} {et~al.}(2019a){Sasikumar Raja}, {Subramanian},
  {Ingale}, \& {Ramesh}}]{Sas2019a}
{Sasikumar Raja}, K., {Subramanian}, P., {Ingale}, M., \& {Ramesh}, R. 2019a,
  \apj, 872, 77

\bibitem[{{Sasikumar Raja} {et~al.}(2017){Sasikumar Raja}, {Subramanian},
  {Ramesh}, {Vourlidas}, \& {Ingale}}]{Sas2017}
{Sasikumar Raja}, K., {Subramanian}, P., {Ramesh}, R., {Vourlidas}, A., \&
  {Ingale}, M. 2017, \apj, 850, 129

\bibitem[{{Sastry} \& {Subramanian}(1974)}]{Sas1974}
{Sastry}, C.~V., \& {Subramanian}, K.~R. 1974, Ind. J. Radio and Space Phys.,
  3, 196

\bibitem[{{Slee}(1959)}]{Sle1959}
{Slee}, O.~B. 1959, Aust. J. Phys, 12, 134

\bibitem[{{Smith} {et~al.}(2001){Smith}, {Mullan}, {Ness}, {Skoug}, \&
  {Steinberg}}]{Smi2001}
{Smith}, C.~W., {Mullan}, D.~J., {Ness}, N.~F., {Skoug}, R.~M., \& {Steinberg},
  J. 2001, \jgr, 106, 18625

\bibitem[{{Spangler}(2002)}]{Spa02}
{Spangler}, S.~R. 2002, The Astrophysical Journal, 576, 997

\bibitem[{{Subramanian}(2000)}]{Subramanian2000}
{Subramanian}, K.~R. 2000, J. Astrophys. Astron., 21, 421

\bibitem[{{Venzmer} \& {Bothmer}(2018)}]{Ven2018}
{Venzmer}, M.~S., \& {Bothmer}, V. 2018, \aap, 611, A36

\bibitem[{{Verdini} {et~al.}(2010){Verdini}, {Velli}, {Matthaeus}, {Oughton},
  \& {Dmitruk}}]{Ver2010}
{Verdini}, A., {Velli}, M., {Matthaeus}, W.~H., {Oughton}, S., \& {Dmitruk}, P.
  2010, \apjl, 708, L116

\bibitem[{{Williams}(1995)}]{Wil1995}
{Williams}, L.~L. 1995, The Astrophysical Journal, 453, 953

\bibitem[{{Woolsey} \& {Cranmer}(2014)}]{Woo2014}
{Woolsey}, L.~N., \& {Cranmer}, S.~R. 2014, \apj, 787, 160

\bibitem[{{Yamauchi} {et~al.}(1998){Yamauchi}, {Tokumaru}, {Kojima},
  {Manoharan}, \& {Esser}}]{Yam98}
{Yamauchi}, Y., {Tokumaru}, M., {Kojima}, M., {Manoharan}, P.~K., \& {Esser},
  R. 1998, Journal of Geophysical Research (Space Physics), 103, 6571

\bibitem[{{Zank} {et~al.}(2018){Zank}, {Adhikari}, {Hunana}, {Tiwari}, {Moore},
  {Shiota}, {Bruno}, \& {Telloni}}]{Zan2018}
{Zank}, G.~P., {Adhikari}, L., {Hunana}, P., {et~al.} 2018, \apj, 854, 32

\end{thebibliography}

%\appendix
\end{document}